\documentclass{sig-alternate-per}

\begin{document}

\title{Cost-Effective Broadcast in Cellular Networks}

\author{
\alignauthor Seung Min Yu and Seong-Lyun Kim\\
       \affaddr{School of Electrical and Electronic Engineering, Yonsei
       University}\\
       \affaddr{50 Yonsei-Ro, Seodaemun-Gu, Seoul 120-749, Korea}\\
       \email{\{smyu,slkim\}@ramo.yonsei.ac.kr}
}

\maketitle
\begin{abstract}
In recent years, video-related services such as YouTube and Netflix
have generated huge amounts of traffic and the network neutrality
debate has emerged as a major issue. In this paper, we consider
feasibility of using a hybrid of unicast and broadcast in cellular
networks for high-traffic services (e.g., video streaming), from the
perspective of cost effectiveness. To reflect spatial
characteristics of base stations (BSs) and mobile users (MUs), we
use the stochastic geometry approach where BSs and MUs are modeled
as independent homogeneous Poisson point processes (PPPs). With
these assumptions and results, we show how to cope with the
trade-off between broadcast and unicast for providing the service
with affordable cost levels and reduced network load. Moreover, we
propose the so called {\it periodic broadcasting service}, where
popular video contents are periodically broadcast over cellular
networks. This service will make a positive impact on the network
neutrality debate by stimulating cooperation between mobile network
operators (MNOs) and content providers (CPs).
\end{abstract}

\category{C.2.1}{Computer-Communication Networks}{Network
Architecture and Design}[Wireless communication]

\terms{Design, Economics}

\section{Introduction}
A Cisco report predicts that global mobile data traffic will
increase 26-fold between 2010 and 2015 \cite{Cisco}. This traffic
explosion drives mobile network operators (MNOs) to invest more in
their networks (upgrading infrastructure and purchasing more
spectrum). On the other hand, MNOs suffer from the decrease of
average revenue per user (ARPU) and severe financial problems. For
this reason, MNOs insist that content providers (CPs) should
shoulder the extra cost for their huge amounts of traffic, which
gives rise to the {\it network neutrality} debate.

Video-related services such as YouTube and Netflix generate huge
amounts of traffic, which already surpassed non-video traffic in
2010 \cite{Cisco}. Naver, a major Internet portal site in South
Korea, provides the live broadcasting service of Korean base ball
game. At the beginning, the service was offered through any access
network, including both wired and wireless. Recently, however, the
service over cellular networks has been terminated due to its huge
amounts of traffic and service quality degradation.

The 3rd Generation Partnership Project (3GPP) has introduced the
Multimedia Broadcast/Multicast Service (MBMS) \cite{3GPP}. An
interesting issue is to find an efficient method for supporting the
MBMS. Broadcast (or multicast) in cellular networks can be a cost
effective way to deliver information to all interested users, by
allowing radio resources to be shared. On the other hand, if the
interested users are few, then it will become a wasteful use of
radio resources. This trade-off is the motivation and a starting
point of our study.

In this paper, we consider a hybrid of unicast and broadcast in
cellular networks for high-traffic services (e.g., video streaming),
from the perspective of cost effectiveness. The cost effectiveness
depends on spatial characteristics of base stations (BSs) and mobile
users (MUs). For reflecting this, we use the stochastic geometry
approach, where BSs and MUs can be modeled as independent
homogeneous Poisson point processes (PPPs) \cite{Andrews}-\cite{Yu}.
This PPP modeling for cellular networks has been strengthened
through theoretical and empirical validation in \cite{Andrews},
\cite{Blaszczyszyn}.\footnote{Many previous studies on cellular
networks assumed that BSs are positioned regularly. However, this
regular model tends to overestimate the performance of cellular
networks due to the perfect geometry of BSs and the neglect of weak
interference from outer tier BSs. For this reason, the PPP modeling
for cellular networks has recently been suggested in
\cite{Andrews}.}

Using a distribution of the number of MUs per cell \cite{Yu}, we
derive some useful metrics for quantifying the wasteful use of radio
resources in unicast and broadcast respectively. With these results,
we evaluate the economic feasibility of broadcast in cellular
networks for video streaming services from cost effectiveness
perspective. Besides, we propose periodic broadcasting service in
cellular networks to increase efficiency, where popular video
contents are periodically broadcast over cellular networks. This
service will make a positive impact on the network neutrality debate
by stimulating cooperation between MNOs and CPs.

\section{System Model}
\begin{figure}
\centering
  \makebox[2.8in]{
        \psfig{figure=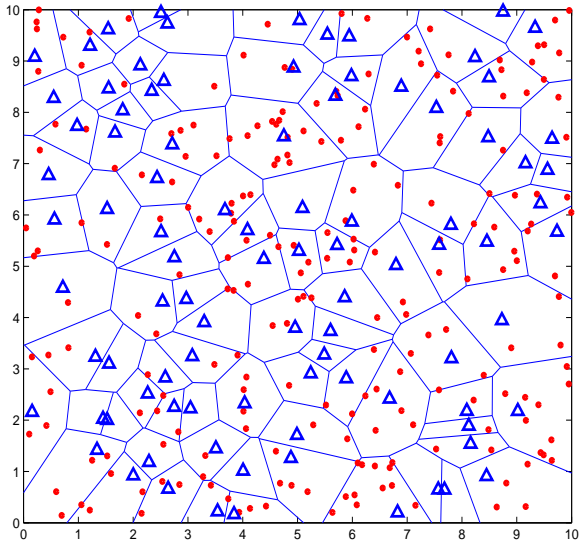,width=3.7in,height=3.3in,clip=;}
  }
\caption{Cellular network where BSs and MUs are distributed as
independent PPPs with $\lambda_u = 3\lambda_b$. Blue triangles and
red dots denote BSs' and MUs' locations, respectively. The cell area
of each BS forms a Voronoi tessellation and the border lines among
Voronoi cells are denoted by blue lines.} \label{system_model}
\end{figure}

Consider a cellular network where BSs and MUs are distributed as
independent PPPs $\Phi_b$ and $\Phi_u$ with density $\lambda_b$ and
$\lambda_u$, respectively. Each MU is assigned to the nearest BS.
Then, the cell area (i.e., coverage) of each BS forms a {\it Voronoi
tessellation} \cite {Okabe} as in Figure \ref{system_model}. A video
content will be streamed over the cellular network. We assume that
the audience rating (i.e., popularity) of the content is $\alpha \in
[0,1]$, and the process of MUs subscribing to the content is an
independent thinning of $\Phi_u$ with the thinning probability
$\alpha$. The value $\alpha$ is close to 1 when the the content is
very popular and a lot of MUs subscribe to it. In the other extreme
($\alpha = 0$), all MUs do not subscribe to the content.

There are two types of video streaming services; {\it buffered video
streaming service} and {\it live video streaming service}. In the
buffered video streaming service such as YouTube and Netflix,
already-produced video contents are streamed over the cellular
network. On the other hand, in the live video streaming service,
video contents are live generated and streamed over the cellular
network. The live video streaming service is synchronized among its
subscribers but the buffered video streaming service is not. In the
next section, we focus on the live video streaming service and
mathematically analyze the trade-off between unicast and broadcast.

\section{Trade-off between Unicast and Broadcast in Cellular Networks}
Assume that MUs can be offered the live video streaming services
through unicast or broadcast in the cellular network. In the unicast
mode, MUs are independently served by using each radio resource even
though they subscribe to the same video content at the same time. On
the other hand, in the broadcast mode, a live video content is
shared by using one radio resource, where all interested MUs can
subscribe to the content simultaneously. If $k$ MUs subscribe to the
content in a cell, then $k-1$ radio resources will be saved by
broadcasting. However, if there is no subscriber (i.e., $k=0$) in
the cell, then one radio resource will go to waste. We denote the
average numbers of saved and wasted radio resources per cell by
$\bar N_s$ and $\bar N_w$, respectively.

To derive $\bar N_s$ and $\bar N_w$, we start with the probability
density function ($f_X(x)$) of the size of a typical Voronoi cell
\cite {Ferenc}:
\begin{eqnarray} \label{eq:cell_size}
f_X \left( x \right) = \frac{{3.5^{3.5} }}{{\Gamma \left( {3.5}
\right)}}\lambda _b^{3.5} x^{2.5} e^{ - 3.5\lambda _b x},
\end{eqnarray}
where $X$ denotes the size of a typical Voronoi cell and its average
value is $E[X]=1/\lambda_b$. Using the distribution of the number of
MUs per a typical Voronoi cell \cite{Yu}, we get a useful
probability mass function in the following:
\begin{eqnarray} \label{eq:pmf_MU}
P\left[ {K = k} \right] = \frac{{3.5^{3.5} \Gamma \left( {k + 3.5}
\right)\left( {\alpha \lambda _u /\lambda _b } \right)^k }}{{\Gamma
\left( {3.5} \right)k!\left( {\alpha \lambda _u /\lambda _b  + 3.5}
\right)^{k + 3.5} }},
\end{eqnarray}
where K denotes the number of MUs who subscribe to a live video
content with the audience rating $\alpha$ in a typical Voronoi cell.
With the law of total probability, we can calculate $E[K]$ as
follows:
\begin{eqnarray} \label{eq:average_MU}
E\left[ K \right] = E\left[ {E\left[ {K|X} \right]} \right] =
E\left[ {\alpha \lambda _u X} \right] = \frac{{\alpha \lambda _u
}}{{\lambda _b }}.
\end{eqnarray}
Then, we can derive $\bar N_s$ and $\bar N_w$ in the following
propositions:

\vskip 10pt \noindent {\bf Proposition 1}: {\it If a live video
content with the audience rating $\alpha$ is broadcast in the
cellular network, then the average number ($\bar N_s$) of saved
radio resources in a typical Voronoi cell is
\begin{eqnarray}
\bar N_s  = \frac{{\alpha \lambda _u }}{{\lambda _b }} + \left( {1 +
3.5^{ - 1} \alpha \lambda _u /\lambda _b } \right)^{ - 3.5} - 1.
\nonumber
\end{eqnarray}
}

\begin{proof}
If there are $K$ MUs subscribing to the content in a typical Voronoi
cell, then $k-1$ radio resources will be saved by broadcasting.
Therefore, we can get the following
equation:\setlength\arraycolsep{1pt}
\begin{eqnarray}
\bar N_s  &=& \sum\limits_{k = 2}^\infty  {\left( {k - 1} \right)P\left[ {K = k} \right]} \nonumber \\
&=& \sum\limits_{k = 2}^\infty  {kP\left[ {K = k} \right]}  - \sum\limits_{k = 2}^\infty  {P\left[ {K = k} \right]} \nonumber \\
&=& \left( {E\left[ K \right] - P\left[ {K = 1} \right]} \right) - \left( {1 - P\left[ {K = 0} \right] - P\left[ {K = 1} \right]} \right) \nonumber \\
&=& E\left[ K \right] + P\left[ {K = 0} \right] - 1  \nonumber \\
&=& \frac{{\alpha \lambda _u }}{{\lambda _b }} + \left( {1 + 3.5^{ -
1} \alpha \lambda _u /\lambda _b } \right)^{ - 3.5} - 1 . \nonumber
\end{eqnarray}
\end{proof}

\vskip 10pt \noindent {\bf Proposition 2}: {\it If a live video
content with the audience rating $\alpha$ is broadcast in the
cellular network, then the average number ($\bar N_w$) of wasted
radio resources in a typical Voronoi cell is
\begin{eqnarray}
\bar N_w  = \left( {1 + 3.5^{ - 1} \alpha \lambda _u /\lambda _b }
\right)^{ - 3.5}. \nonumber
\end{eqnarray}
}

\begin{proof}
If there is no subscriber in a typical Voronoi cell (i.e., $K=0$),
then one radio resource will be wasted by broadcasting. Therefore,
we can get the following equation:\setlength\arraycolsep{1pt}
\begin{eqnarray}
\bar N_w  = 1 \times P\left[ {K = 0} \right] = \left( {1 + 3.5^{ -
1} \alpha \lambda _u /\lambda _b } \right)^{ - 3.5}. \nonumber
\end{eqnarray}
\end{proof}

\begin{figure}[t]
\centerline{\epsfig{figure=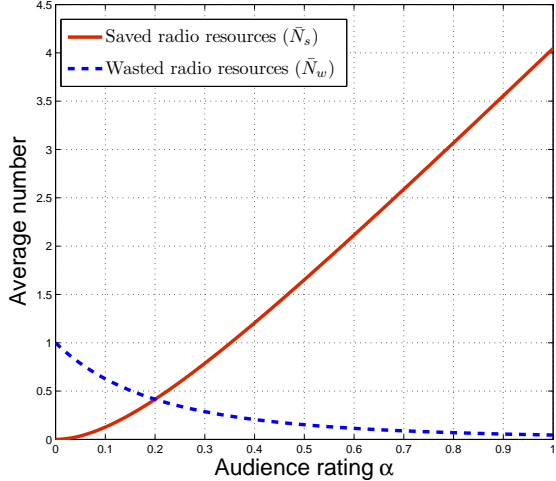,height=2.8in,clip=;}}
\caption{Average numbers of saved ($\bar N_s$) and wasted ($\bar
N_w$) radio resources in a typical Voronoi cell as a function of the
audience rating $\alpha$. Here, we fix $\lambda_u / \lambda_b$ to be
3.} \label{analysis_broadcast_figure}
\end{figure}

\noindent Propositions 1 and 2 show that if the MU density is much
higher than the BS density like downtowns (i.e., $\lambda_u \ \gg
\lambda_b$), then $\bar N_s$ and $\bar N_w$ can be approximated to
$\alpha \lambda_u / \lambda_b -1$ and zero, respectively. On the
other hand, if the live video content is very unpopular (i.e.,
$\alpha \approx 0$) or the BS density is much higher than the MU
density like femtocells (i.e., $\lambda_b \ \gg \lambda_u$), then
$\bar N_s$ and $\bar N_w$ can be approximated to zero and one,
respectively. This means that broadcast is more desirable in the
former case but unicast is better in the latter case (i.e.,
trade-off between unicast and broadcast), which is also shown in
Figure \ref{analysis_broadcast_figure}.

\section{Cost Effectiveness of Broadcast in Cellular Networks}
In this section, we show cost effectiveness of broadcasting a live
video content over the cellular network. For this, we define the
average amount of cost reduction ($CR$) by broadcast as follow:
\begin{eqnarray} \label{eq:cost_reduction_definition}
CR = v_r \bar N_s  - v_r \bar N_w  - c_b,
\end{eqnarray}
where $v_r$ and $c_b$ denote the monetary value of unit radio
resource and the additional cost for broadcast implementation,
respectively. In Equation (\ref{eq:cost_reduction_definition}), $v_r
\bar N_s$ and $v_r \bar N_w$ represent the saved and wasted costs by
broadcast, respectively. With Propositions 1 and 2, we can get the
following proposition.

\vskip 10pt \noindent {\bf Proposition 3}: {\it If the audience
rating $\alpha$ of a live video content satisfies the following
condition;
\begin{eqnarray}
\alpha  > \frac{{\lambda _b }}{{\lambda _u }}\left( {\frac{{c_b
}}{{v_r }} + 1} \right), \nonumber
\end{eqnarray}
then the average amount of cost reduction ($CR$) by broadcasting the
content over the cellular network is;
\begin{eqnarray}
CR = v_r \left( {\frac{{\alpha \lambda _u }}{{\lambda _b }} - 1}
\right) - c_b. \nonumber
\end{eqnarray}
}

\begin{proof}
By broadcasting a live video content over the cellular network, the
average value $v_r \bar N_s$ is saved but the average value $v_r
\bar N_w$ is wasted. Moreover, there is an additional cost $c_b$.
With Propositions 1 and 2, the average amount of cost reduction $CR$
can be calculated as follows:
\begin{eqnarray} \label{eq:cost_reduction}
CR = v_r \bar N_s  - v_r \bar N_w  - c_b  = v_r \left({\frac{{\alpha
\lambda _u }}{{\lambda _b }} - 1} \right) - c_b.
\end{eqnarray}
By re-arranging Equation (\ref{eq:cost_reduction}), we can find the
audience rating condition for making $CR$ positive.
\end{proof}

\vskip 10pt Considering the value of radio resources increased in
these days, Proposition 3 show that the cost reduction gain by
broadcasting over the cellular network will be high in practice.
Proposition 3 is derived under an implicit assumption that all MUs
subscribing to a live video content will immediately switch to the
broadcasting channel. In practice however, some MUs might prefer
unicast if there is no economic incentive in broadcast, and the
average amount of cost reduction (Proposition 3) will decrease. In
this case, the service price discount on broadcast may be an
attractive option for the MUs. Proposition 3 is not applicable to
the buffered video streaming service because there will be less MUs
subscribing to the same buffered video content concurrently. More
details are described in the next section.

\section{Periodic Broadcasting Service for Buffered Video Contents}
In the buffered video streaming service, already-produced video
contents are streamed over the cellular network and MUs can
subscribe to the contents whenever they want. This means that
broadcasting the content is very inefficient. To tackle this
inefficiency, we propose {\it periodic broadcasting service for
buffered video contents}, where some buffered video contents are
periodically broadcast in the cellular network. If economic
incentives such as price discount are offered, the service can bring
a cost reduction effect.

\begin{figure}[t]
\centerline{\epsfig{figure=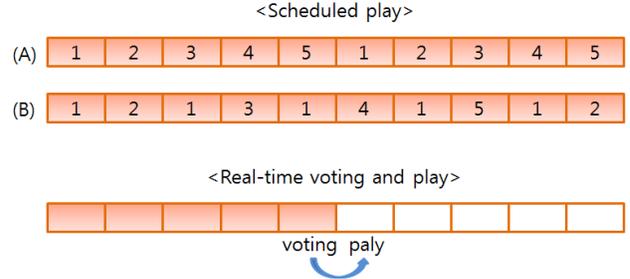,height=1.5in,clip=;}}
\caption{Scheduled play, and real-time voting and play.}
\label{broadcasting_scheme}
\end{figure}

In this service, it is very important to determine what and how
video contents are played. For this, we suggest two simple schemes;
{\it scheduled play} and {\it real-time voting and play}. In the
scheduled play scheme, video contents are sorted by the popularity
(e.g., the cumulative number of views on YouTube or Netflix)
\cite{Cha} and top-$n$ video contents are played repeatedly. Figure
\ref{broadcasting_scheme} shows two examples of the scheduled play
scheme, where the top-5 video contents are played with an equal
weight (A) and with a different weight (B). The scheduled play
scheme is simple but it cannot reflect the real-time changing of
user demand. On the other hand, in the real-time voting and play
scheme, MUs vote for their favorite contents during a certain period
and a video content receiving the most votes is played in the next
period (Figure \ref{broadcasting_scheme}).

For the success of periodic broadcasting services, MNOs can go into
business in partnership with CPs. These services can reduce
inefficiency in the video content delivering and can create
additional revenues by advertising, which will eventually give a
positive effect on the network neutrality debate.

\section{Conclusions}
In this paper, we consider broadcast in cellular networks for video
streaming services and analyze its cost effectiveness reflecting
spatial characteristics of BSs and MUs. Using the stochastic
geometry approach, we derive average numbers of saved and wasted
radio resources in a typical Voronoi cell respectively, and show
there is the trade-off between broadcast and unicast. With these
results, we evaluate the economic feasibility of broadcast in
cellular networks for live video streaming services. Moreover, we
propose periodic broadcasting service for buffered video contents.
Even though our analytic results are derived under a simple model
for mathematical tractability, it will provide engineering and
economic insights for managing huge amounts of video traffic in
cellular networks.

In recent years, the network neutrality debate has been emerging as
a major issue. Moreover, the mobile data explosion causes the
spectrum shortage and the data usage polarization among users, which
will eventually lead to the decrease of user welfare. For this
reason, in our previous work \cite{Wpin}, we have proposed a data
subsidy scheme where the regulator offers spectrum price discount to
MNOs in return for providing a predefined data amount to users
without any charge, and have showed that it can increase user
welfare even further without MNOs' profit loss. Broadcasting service
in cellular networks for live and buffered video contents can be
directly applied to the data subsidy, which will generate a
significant synergy effect on both solving the mobile data explosion
and improving user welfare.

\end{document}